\documentclass[aps,prl,twoside,twocolumn,10pt,floatfix,showpacs,citeautoscript,superscriptaddress]{revtex4-2}

\usepackage{amsmath}
\usepackage{amssymb}
\usepackage{booktabs}
\usepackage{comment}
\usepackage{glossaries}
\usepackage{graphicx}
\usepackage[version=4]{mhchem}
\usepackage{siunitx}
\usepackage{tabularx}
\usepackage[dvipsnames]{xcolor}

\usepackage[colorlinks=true]{hyperref}

\hypersetup{
  colorlinks,
  citecolor=black,
  linkcolor=black,
  urlcolor=black}

\hypersetup{pdfauthor={Fredrik Erikssonm Yulong Qiao, Erik Fransson, R. Matthias Geilhufe, and Paul Erhart}}
\hypersetup{pdftitle={Non-Markovian heat production in ultrafast phonon dynamics}}

\graphicspath{{figures/}}

\setacronymstyle{long-short}
\newacronym{dft}{DFT}{density functional theory}
\newacronym{dos}{DOS}{density of states}
\newacronym{md}{MD}{molecular dynamics}
\newacronym{mlip}{MLIP}{machine-learned interatomic potential}
\newacronym{nep}{NEP}{neuroevolution potential}
\newacronym{psd}{PSD}{power spectral density}
\newacronym{sto}{STO}{\ce{SrTiO3}}

\makeatletter
\let\oldtheequation\theequation
\renewcommand\tagform@[1]{\maketag@@@{\ignorespaces#1\unskip\@@italiccorr}}
\renewcommand\theequation{(\oldtheequation)}
\makeatother

\DeclareSIUnit\angstrom{\text{Å}}
\renewcommand{\vec}[1]{\mathbf{#1}}
\renewcommand{\epsilon}[0]{\varepsilon}

\newcommand{\addchalmers}{
    Department of Physics,
    Chalmers University of Technology,
    SE-41296, Gothenburg, Sweden
}

\begin{document}

\title{Non-Markovian heat production in ultrafast phonon dynamics}
\author{Fredrik Eriksson}
\author{Yulong Qiao}
\email{yulong.qiao@chalmers.se}
\author{Erik Fransson}
\author{R. Matthias Geilhufe}
\email{matthias.geilhufe@chalmers.se}
\author{Paul Erhart}
\email{erhart@chalmers.se}
\affiliation{\addchalmers}

\date{\today}

\begin{abstract}
High-intensity THz laser pulses enable the light-mediated control of lattice vibrations by resonantly driving selected phonon modes.
On ultrafast timescales, memory effects influence the phonon dynamics and must be accounted for to describe the heat production associated with energy dissipation.
Here, we establish a microscopic framework for non-Markovian phonon dynamics by deriving the noise and dissipation kernels governing a driven phonon mode.
Using large-scale molecular dynamics simulations, we reconstruct these kernels directly from the many-body lattice dynamics and determine the corresponding heat production rate.
Our results provide a quantitative picture of the crossover between Markovian and non-Markovian dynamics on picosecond timescales and show how the finite bandwidth of the driving field limits the dynamically relevant bath spectrum.
Furthermore, we demonstrate that thermodynamic quantities such as heat production can be inferred directly from the dynamics of an individual phonon mode, enabling their experimental measurement using time-resolved spectroscopy.
\end{abstract}

\maketitle

Intense THz laser pulses enable the selective excitation and coherent control of optical phonon modes \cite{Disa2021, Salen2019, Nicoletti2016}.
Driving the lattice along specific vibrational coordinates has opened nonthermal pathways to metastable phases \cite{Fausti2011, Subedi2017, Nova2019, Buzzi2021, Budden2021, Latini2021, Cheng2023}, phonon-induced magnetism \cite{Juraschek2017, Juraschek2019, Cheng2020, Geilhufe2021, Ren2021, Juraschek2022, Baydin2022, Basini2022, Geilhufe2023, Luo2023, Hernandez2023, Shabala2024, Shabala2025}, and ultrafast magnetic switching \cite{Disa2020, Davies2023}.
These advances raise a fundamental question: how is heat produced when a solid is driven far from equilibrium on picosecond timescales?

Recent theoretical work has begun to frame laser-driven phonons within stochastic thermodynamics, highlighting the interplay of fluctuations and dissipation in generating heat and entropy \cite{Caprini2024, Qiao2025, Tietjen2025}.
In such descriptions, the driven mode acts as a subsystem coupled to an effective bath formed by the remaining lattice degrees of freedom.
Energy transfer from coherent motion into this bath enhances fluctuations and results in entropy production.
However, the microscopic structure of this bath and the validity of Markovian assumptions under strong driving remain largely unexplored.

Here we establish a microscopic framework for ultrafast thermodynamics that resolves heat production at the level of individual phonon modes.
By combining large-scale \gls{md} simulations with a mode-resolved theoretical description, we obtain deterministic access to the full many-body lattice dynamics and project it onto normal-mode coordinates.
This approach enables a direct and quantitative comparison between atomistic simulations and stochastic effective theories.
From this comparison we derive a non-Markovian Langevin equation for a laser-driven phonon mode and identify the microscopic origin of memory effects in terms of mode–mode coupling.

We show that far-from-equilibrium driving generates a highly structured effective bath whose spectral density is dominated by discrete phonon modes rather than a continuum.
Despite this structure, a Markovian approximation can remain quantitatively accurate once the finite frequency resolution of the driving field is taken into account.
Our results clarify when simplified stochastic descriptions capture ultrafast heat production and when genuinely non-Markovian effects become essential.

\begin{figure*}
\centering
\includegraphics[height=1.9in]{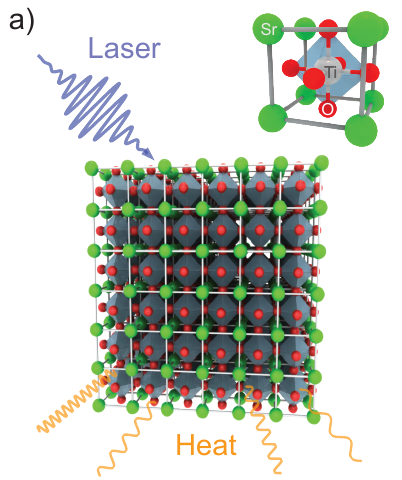}%
\includegraphics[height=2.1in]{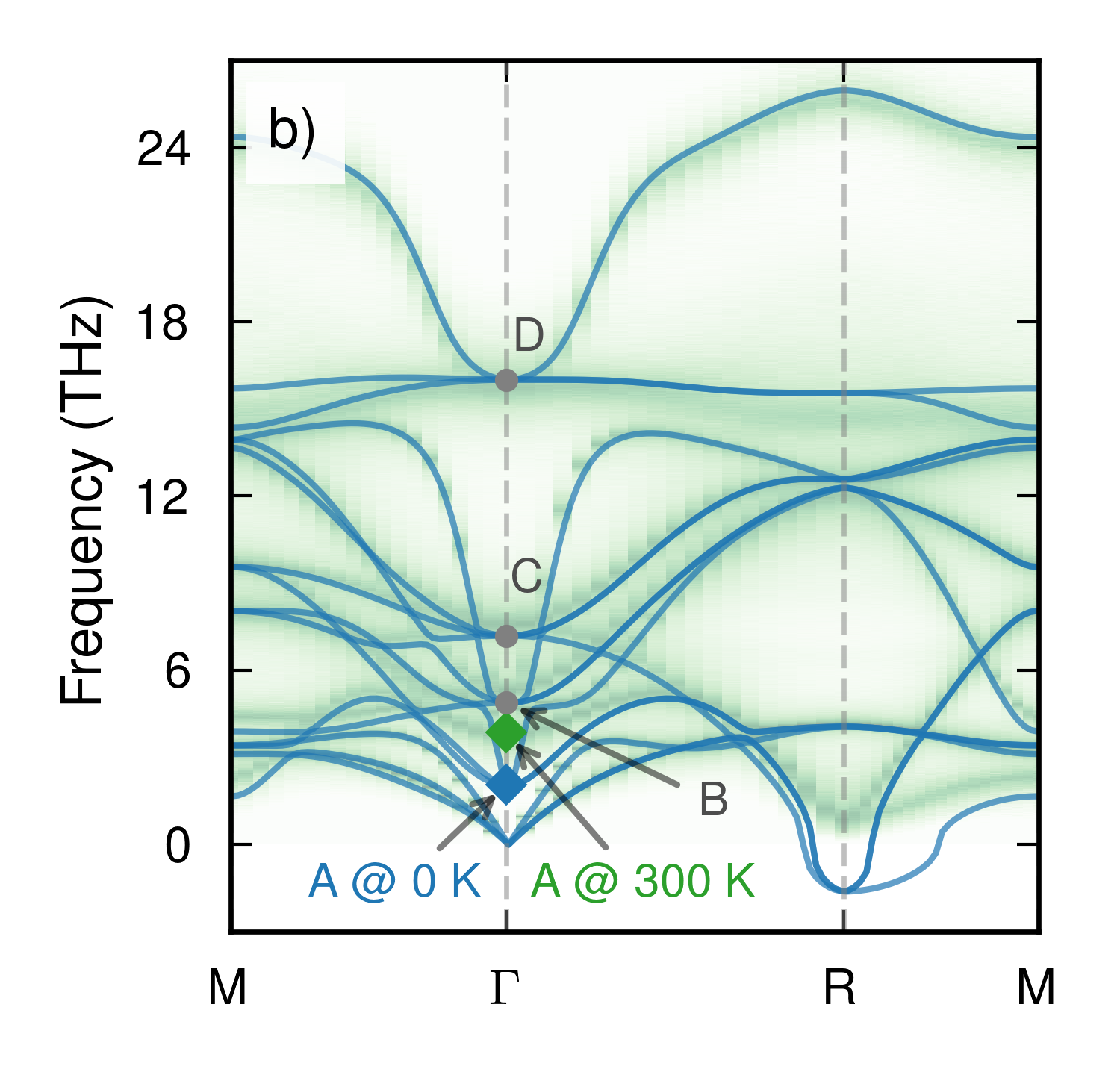}%
\includegraphics[height=2.1in]{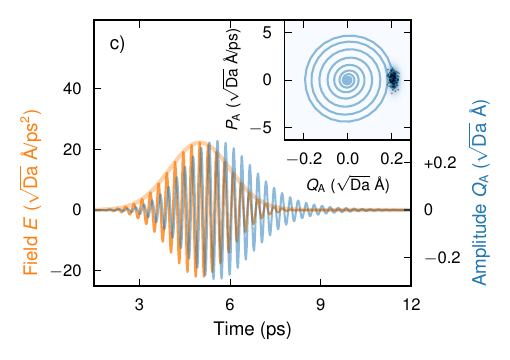}
\caption{
    (a) Schematic illustrating of \gls{sto} driven by a THz laser pulse, leading to pumping of the ferroelectric mode, which on decay leads to the ultrafast generation of heat.
    The inset shows the primitive unit cell.
    (b) Phonon band structure along a high-symmetry path through the Brillouin zone.
    Solid blue lines show the harmonic dispersion corresponding to the zero-temperature limit.
    The green color map shows the spectral energy density representing the dispersion at \qty{300}{\kelvin}.
    The ferroelectric mode (marked A), has a pronounced temperature dependence.
    The other $\Gamma$-modes are labeled B through D (see \autoref{sfig:temperature-dependence-gamma-modes} for a more explicit illustration of their temperature dependence).
    (c) Excitation field due to the laser (in orange; with envelope indicated) and the mode amplitude of the ferroelectric mode (in blue) as a function of time.
    The inset shows the sampling of the phase space of the driven mode as a kernel-density map over the distribution of trajectories with markers representing specific simulations at approximately \qty{5.59}{\pico\second} corresponding to the maximum amplitude.
    The line shows the average path through phase space up to this time.
}
\label{fig:simulation-overview}
\end{figure*}

As a concrete realization we consider the ferroelectric soft mode of the perovskite \gls{sto}, a prototypical quantum paraelectric with strongly anharmonic lattice dynamics \cite{Kozina2019, Latini2021, Basini2022, Caprini2024}.
Its proximity to structural instability and pronounced temperature-dependent softening make it an ideal platform for exploring nonequilibrium phonon thermodynamics (\autoref{snote:sto}) \cite{Fleury1968, Cowley1996, Holt2007, muller1979srti, VerRanFra23}.
While the material serves as a representative example, the framework developed here is general and applicable to driven lattice systems more broadly.

To capture the interplay of lattice instabilities, fluctuations, and dissipation beyond harmonic models, we employ a \gls{mlip} trained within the \gls{nep} framework \cite{FanWanYin22, xu2025mega, LinRahFra24, schaul2011high} as an efficient emulator of \gls{dft} calculations \cite{KreFur1996-1, Blo94, KreJou99} using the van-der-Waals density functional with consistent exchange \cite{DioRydSch04, BerHyl2014}.
This description accurately reproduces the relevant lattice dynamics and anharmonic couplings governing energy redistribution.
To investigate the nonequilibrium dynamics of the ferroelectric soft mode following optical excitation, and to analyze the associated dissipation mechanisms and thermodynamic changes, we performed a large ensemble of \gls{mlip}-\gls{md} simulations.
In these simulations, the laser pulse was modeled as an effective time-dependent driving force $F_\text{L}(t)$ acting directly on the soft-mode amplitude,
\begin{align}
    F_\text{L}(t) = \beta Z E \exp\left(-\frac{1}{2} \frac{(t-t_0)^2}{\tau^2}\right) \cos\left(\omega_\text{L} t\right).
    \label{eq:laser}
\end{align}
Here, $Z$ denotes the mode effective charge \cite{Gonze1997}, which is $Z \approx \qty{1.2}{e^+/\sqrt{\dalton}}$ (see \autoref{snote:mode-effective-charge}), and $\beta$ accounts for the screening of the electric field within the sample, approximated as $\beta = 0.215$ \cite{Kozina2019}.
We used a force amplitude of \qty{22.2}{\sqrt{\dalton}\angstrom\per\pico\second\squared}, corresponding to an electric field of $E \approx \qty{860}{\kilo\volt\per\centi\meter}$, comparable to values employed in recent experiments \cite{Salen2019, Kozina2019, Luo2023, Basini2022, Davies2023}.
The pulse duration was set to $\tau=\qty{1}{\pico\second}$ with its peak centered at $t_0=\qty{5}{\pico\second}$ (\autoref{fig:simulation-overview}c), and the central frequency $\omega\approx \qty{3.66}{\tera\hertz}$ was chosen close to resonance with the computed room-temperature soft-mode frequency of approximately \qty{3.84}{\tera\hertz} (\autoref{stab:irreps}).

We obtain the time-dependent mode coordinates $Q_{\lambda}(t)$ for a phonon mode $\lambda=(\vec{k},n)$ by projecting each \gls{md} configuration onto the corresponding phonon eigenvectors \cite{SunSheAll2010, Rohskopf2022, BerFraEri25}.
For the ferroelectric soft mode at the $\Gamma$ point, $Q_\text{A}=Q_{\vec{0},1}$, this projection yields
\begin{align}
    Q_\text{A}=\frac{1}{\sqrt{N_c}}\sum_{\nu,\alpha}\sqrt{m_{\nu}}\mathbf{e}^A_{\nu\alpha}\tau^A_{\nu\alpha},
    \label{eq:u0}
\end{align}
where $N_c$ is the number of unit cells and $\mathbf{e}^A_{\nu\alpha}$ denotes the normalized eigenvector of the soft mode at $\Gamma$, with $\nu$ and $\alpha$ labeling atomic species and Cartesian direction, respectively.
Here, $m_{\nu}$ is the atomic mass and $\tau^A_{\nu\alpha}$ the Fourier component of the atomic displacement at the $\Gamma$-point.
The mass-weighted mode coordinate $Q_\text{A}$ thus has units of \unit{\sqrt{\dalton}\,\angstrom}.

\begin{figure*}
    \centering
    \includegraphics[height=2.1in]{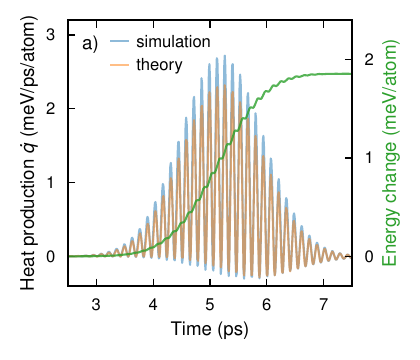}
    \includegraphics[height=2.1in]{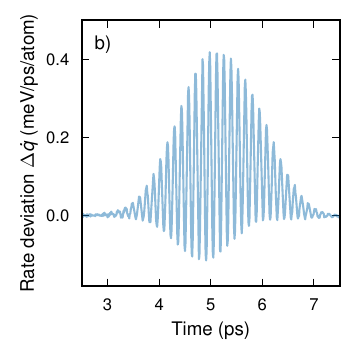}
    \includegraphics[height=2.1in]{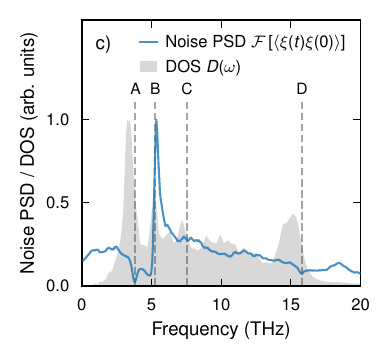}
    \caption{
        (a) Total energy and heat production rate $\dot{q}(t)$ during pulse, where $\dot{q}(t)$ is obtained as the time derivative of the total energy.
        (b) Deviation in heat production rate between simulation and theory.
        (c) Noise \gls{psd} $\mathcal{F}\left[\langle\xi(t)\xi(0)\rangle\right]$ of the ferroelectric mode (A in \autoref{fig:simulation-overview}b) in comparison with the \acrfull{dos}.
        The vertical dashed lines indicate the frequencies of the $\Gamma$-point modes (see \autoref{sfig:noise-acf-modes-and-dos} for the noise \glspl{psd} for all modes).
    }
    \label{fig:heat-production}
\end{figure*}

The laser excitation drives the soft mode to a maximum amplitude of almost \qty{0.3}{\sqrt{\dalton}\angstrom} (blue line in \autoref{fig:simulation-overview}c).
Because the pulse duration (\qty{1}{\pico\second}) exceeds the oscillation period of the mode, several coherent oscillations occur during the driving.
However, the strong damping of the ferroelectric soft mode limits its lifetime to a timescale comparable to the pulse duration, and the amplitude rapidly relaxes toward equilibrium once the driving field vanishes.
This pronounced damping reflects the energy transfer from the driven mode to the surrounding lattice degrees of freedom.

To interpret the phonon dynamics and associated energy transfer processes observed in the \gls{md} simulations, we formulate an ultrafast stochastic thermodynamic framework based on a classical phonon--bath coupling model.
A full quantum mechanical treatment is discussed in Ref.~\cite{Qiao2025}.
Here we focus on the classical limit appropriate for the large-amplitude, finite-temperature dynamics considered in this work.
We partition the phonon system into the driven ferroelectric soft mode and an effective bath comprising all remaining lattice degrees of freedom, and describe their coupled dynamics by the classical Lagrangian \cite{Kemeny1986},
\begin{multline}
    \mathcal{L} = \frac{\dot{Q}_\text{A}^2}{2}
    - \frac{1}{2} \omega_\text{A}^2 Q_\text{A}^2 + F_\text{L}(t) Q_\text{A} 
    \\ + \sum_{i=1}\left[\frac{\dot{Q}_i^2}{2} - \frac{\omega_i^2}{2}\left(Q_i-\frac{\kappa_i}{\omega_i^2}Q_\text{A}\right)^2\right].
    \label{eq:la}
\end{multline}
Here, $Q_\text{A}(t)$ denotes the amplitude of the ferroelectric mode defined in \autoref{eq:u0}, $\dot{Q}_\text{A}(t)$ its generalized velocity, and $\omega_\text{A}$ its angular frequency.
The laser force $F_\text{L}(t)$ is given by \autoref{eq:laser}.
The bath is represented by a large set of harmonic modes indexed by $i$.
Although real phonon--phonon interactions involve anharmonic couplings of various polynomial orders, a linear mode--bath coupling term $\kappa_i Q_\text{A} Q_i$ provides an effective description of dissipation and energy transfer from the soft mode at the level of coarse-grained dynamics.

Eliminating the bath degrees of freedom yields the effective equation of motion for the soft mode (see \autoref{snote:microscopic-model} for details),
\begin{align}
    \Ddot{Q}_\text{A}+\omega_\text{A}^2Q_\text{A}+\int_{-\infty}^t\mathrm{d}s\, f(t-s) \, \dot{Q}_\text{A}(s)=F_\text{L}(t)+\xi_\text{A}(t),
    \label{eq:non-markov}
\end{align}
which takes the form of a generalized Langevin equation.
Dissipation is governed by the memory kernel $f(t)$ through the convolution term, where
\begin{equation}
     f(t) = \int_{0}^{\infty}\mathrm{d}\omega\, D(\omega)\frac{\kappa^2(\omega)}{\omega^2}\cos(\omega t),
     \label{eq:memory}
\end{equation}
and depends on the phonon density of states $D(\omega)$ and the mode--bath coupling strength $\kappa(\omega)$.
The stochastic force $\xi(t)$ is a Gaussian colored noise with correlation function $\eta(t)=\langle \xi(t)\xi(0)\rangle$ satisfying the classical fluctuation--dissipation theorem in the Fourier domain,
\begin{equation}
    \mathcal{F}\left[\eta(t)\right] = k_{\text{B}}T\,\mathcal{F}\left[f(t)\right],
    \label{eq:FD}
\end{equation}
where $\mathcal{F}[\cdots]$ denotes the Fourier transform.
The quantity $\mathcal{F}\left[\eta(t)\right]$ corresponds to the \acrfull{psd} of the stochastic force and, according to \autoref{eq:memory} and \autoref{eq:FD}, is proportional to $D(\omega)\kappa^2(\omega)/\omega^2$ up to a temperature factor.

The generalized Langevin equation \eqref{eq:non-markov} describes non-Markovian dynamics due to the memory kernel entering through the convolution term.
In practice, however, the kernel $f(t)$ can often not be directly accessed from simulations or experiments.
If $f(t)$ can be approximated as time-local, the equation of motion reduces to
\begin{align}
    \Ddot{Q}_\text{A}
    + \omega_\text{A}^2 Q_\text{A}
    + \gamma_\text{A} \dot{Q}_\text{A}
    = F_\text{L}(t) + \xi_\text{A}(t),
    \label{eq:markov}
\end{align}
where $\gamma_\text{A}$ denotes the damping rate and $\omega_\text{A}$ the renormalized soft-mode frequency.
In this limit, $Q_\text{A}(t)$ follows Markovian dynamics \cite{Caprini2024}.

To extract the stochastic force $\xi_\text{A}(t)$ from the \gls{md} simulations, we project the atomic forces (excluding the laser force $F_\text{L}(t)$) onto the phonon eigenmodes (\autoref{snote:stochastic-force-from-md}).
For the modes at $\Gamma$, including the ferroelectric soft mode~A, we perform a least-squares fit of the total mode force to the form $\omega_\lambda^2 Q_\lambda(t) + \gamma_\lambda \dot{Q}_\lambda(t)$ to determine the renormalized frequencies $\omega_\lambda$ and damping rates $\gamma_\lambda$.
Frequency shifts induced by anharmonic interactions are thus absorbed into $\omega_\lambda$, while energy dissipation into other modes is captured by $\gamma_\lambda$.
The stochastic force $\xi_\lambda(t)$ is identified as the residual after subtracting the harmonic and friction contributions.
After this fitting procedure, the dynamics of the soft mode coordinate ($Q_A(t)$ in \autoref{fig:simulation-overview}c) is fully captured by the time-local Langevin equation \autoref{eq:markov}, upon averaging over the zero-mean stochastic force.
The parameters are given by the soft mode frequency $\omega_A = \qty{3.84}{\tera\hertz}$ and the soft mode damping $\gamma_A = \qty{0.35}{\tera\hertz}$ (\autoref{stab:irreps}).

The increase in the total energy of the system, $\Delta E(t)$, relative to its pre-pulse value can be interpreted as the net heat deposited into the lattice due to the work performed by the laser.
This energy change can be written as an integral over the heat production rate $\dot{q}(t)$,
\begin{equation}
    \Delta E(t) = \int_{-\infty}^{t} \mathrm{d} \tau\, \dot{q}(\tau).
    \label{eq:heat}
\end{equation}
In the \gls{md} simulations, $\Delta E(t)$ is obtained by summing the kinetic and potential energies of all atoms (green line in \autoref{fig:heat-production}a).
The total energy increases rapidly during the pump and saturates shortly after the pulse, indicating that the energy deposition process is essentially completed within \qty{3}{\pico\second}.
The instantaneous heat production rate $\dot{q}(t)$ is computed as the time derivative of the total energy.
It oscillates during the pulse and reaches a maximum near \qty{5}{\pico\second}, consistent with the soft-mode dynamics (blue line in \autoref{fig:simulation-overview}c).
The small negative contribution around \qty{6}{\pico\second} arises from the slight frequency mismatch between the laser and the soft mode.

Within the stochastic thermodynamic description based on on the generalized Langevin equation \eqref{eq:non-markov}, the heat production rate can be evaluated analytically.
At the level of a single trajectory, the time derivative of the internal energy $U_\text{A}(t)$ of the soft mode
is given by
%$U_\text{A}(t)=\tfrac{1}{2}\omega_\text{A}^2Q_\text{A}^2+\tfrac{1}{2}\dot{Q}_\text{A}^2$, which gives for its time derivative
%\begin{align}
 %   \dot{U}_\text{A}(t) &=
 %   F_\text{L}(t)\dot{Q}_\text{A}(t)
 %   + \left[\xi_A(t)
 %   - \gamma_A\dot{Q}_\text{A}(t)\right] \circ \dot{Q}_\text{A},
 %   \label{eq:dU}
%\end{align}
%\begin{multline}
\begin{equation}
    \dot{U}_\text{A}(t) =
    F_\text{L}(t)\dot{Q}_\text{A}(t)
    + \Delta \dot{Q}.
    %\Big[\xi_A(t) \\
    %- \int_{-\infty}^t\mathrm{d}s\, f(t-s) \, \dot{Q}_\text{A}(s)\Big]\circ \dot{Q}_\text{A},
\end{equation}
%\end{multline}
The first term represents the work rate performed by the laser on the soft mode.
The second term quantifies the stochastic energy flux from the thermal bath into the soft mode,
\begin{equation}
    \Delta \dot{Q}=\left[\xi_A(t)
    - \int_{-\infty}^t\mathrm{d}s\, f(t-s) \, \dot{Q}_\text{A}(s)\right]\circ \dot{Q}_\text{A}.
\end{equation}
Here, $\circ$ denotes the Stratonovich product \cite{dabelow2019}.
Taking the ensemble average and using the first-law relation, the heat production rate becomes
\begin{align}
    \dot{q}(t)
    &= \left< \dot{U}_\text{A}(t) - \Delta\dot{Q}(t) \right>
    =
    \left< F_\text{L}(t) \dot{Q}_\text{A}(t) \right>,
    \label{eq:qt}
\end{align}
where $\langle\cdots\rangle$ denotes the ensemble average over stochastic trajectories.
The heat production rate is therefore determined by the correlation between the laser force and the velocity of the driven phonon mode.

The form of \autoref{eq:qt} is general, while non-Markovian effects enter through the dynamics of $Q_A(t)$ governed by the generalized Langevin equation \autoref{eq:non-markov} with memory kernel $f(t)$.
Using the Markovian approximation \autoref{eq:markov} together with \autoref{eq:qt} yields the theoretical prediction shown as the orange line in \autoref{fig:heat-production}a.
Comparison with the \gls{md} result (blue line) shows very good agreement, although the theory slightly overestimates the peak value (\autoref{fig:heat-production}b).
This agreement confirms that the stochastic thermodynamic description captures the essential mechanisms governing energy transfer during ultrafast lattice excitation.

To shed light into the form of the actual memory kernel $f(t)$ we compute the \gls{psd} of the zero-mean stochastic forces using the Fourier transform of the autocorrelation function $\langle\xi(t)\xi(0)\rangle$.
By construction, the \gls{psd} vanishes at the respective eigenfrequency of each mode.
The \glspl{psd} are clearly non-uniform, demonstrating that the effective noise is colored (\autoref{fig:heat-production}c; \autoref{sfig:noise-acf-modes-and-dos}).
Comparison of the \gls{psd} for the ferroelectric soft mode~A with the phonon density of states $D(\omega)$ shows that the spectral structure does not simply follow $D(\omega)$, highlighting the role of the coupling strength $\kappa(\omega)$.
In particular, the \gls{psd} for mode~A exhibits a pronounced peak at $\omega_\text{B}=\qty{5.26}{\tera\hertz}$, revealing strong coupling to mode~B.
Applying the same analysis to the remaining $\Gamma$-point modes shows a much weaker coupling to mode~D and no observable coupling to mode~C, consistent with modes~A and C belonging to different symmetry representations (see \autoref{stab:irreps}).
Similar spectral features have previously been observed in time-resolved X-ray scattering experiments \cite{Kozina2019}.

The structured \gls{psd} indicates a nonlocal friction kernel and thus, in principle, non-Markovian dynamics according to \autoref{eq:FD}.
However, the observed soft-mode dynamics is well captured by the Markovian equation \autoref{eq:markov}.
This apparent discrepancy can be understood by considering the finite frequency resolution of the laser pulse.
The driven response of the mode is governed by the product of the frequency response function $\chi(\omega)$ defined in \autoref{snote:microscopic-model}  and the spectral profile of the driving field $F_\text{L}(\omega)$.
Because the pulse duration $\tau=\qty{1}{\pico\second}$ is much longer than the soft-mode period, the spectrum of the driving field is narrowly peaked around $\omega_\text{L}$.
Within this restricted frequency window, the response function varies only weakly and can be approximated by its local expansion, effectively yielding a constant damping rate $\gamma$.
As a result, the non-Markovian dynamics reduces to an effective Markovian form on the timescale set by the pulse.
% A quantitative illustration based on a Drude–Lorentz spectral density confirms that the resulting dynamics is nearly indistinguishable from the full non-Markovian solution under the present driving conditions (see \autoref{snote:markovian-vs-non-markovian}).
For shorter pulses with broader spectral bandwidth, deviations from Markovian behavior are expected to become observable, as illustrated in \cite{Qiao2025}.

In conclusion, we have developed a microscopic framework to quantify non-Markovian heat production in ultrafast phonon dynamics by integrating large-scale \gls{md} simulations with a stochastic, mode-resolved theoretical description.
Using a \gls{mlip} with near--\gls{dft} accuracy, we resolved the full many-body lattice dynamics of a realistic solid, projected them onto phonon normal modes, and reconstructed the structured effective bath governing a laser-driven phonon mode.
This approach establishes a direct bridge between deterministic atomistic simulations and stochastic thermodynamics, enabling the extraction of memory kernels and bath spectral densities from first principles.

The resulting generalized Langevin description quantitatively reproduces the ultrafast dynamics observed in the simulations and reveals strongly mode-selective energy transfer within the lattice.
We show that the heat production rate can be inferred directly from the driven-mode dynamics, yielding excellent agreement with the fully atomistic simulations.
Our analysis further demonstrates that far-from-equilibrium driving generates a highly structured bath with a non-uniform \gls{psd}, reflecting discrete phonon couplings rather than a featureless continuum.
Nevertheless, the finite frequency bandwidth of the driving pulse limits the dynamically relevant spectral window, such that an effective Markovian description remains accurate on picosecond timescales.

While illustrated here for the ferroelectric soft mode of \gls{sto}, the framework is general and applicable to driven lattice systems beyond this specific material realization.
Our results provide a quantitative and mode-resolved understanding of how dissipation, heat production, and fluctuations emerge in strongly driven solids.
More broadly, this work establishes a route toward first-principles-informed ultrafast thermodynamics, where non-Markovian effects can be identified, quantified, and systematically explored in experiments on quantum materials.

\section*{Data availability}
The \gls{dft} data and \gls{nep} model generated in this study are openly available via Zenodo at \url{https://doi.org/10.5281/zenodo.18841778}.

\section{acknowledgments}
FE, EF, and PE acknowledge funding from the Swedish Research Council (Nos. 2020-04935 and 2025-03999) and the Knut and Alice Wallenberg Foundation (No. 2024.0042).
RMG and YQ acknowledge support from the Swedish Research Council (VR Starting Grant No. 2022-03350), the Olle Engkvist Foundation (No. 229-0443), the Royal Physiographic Society in Lund (Horisont), the Knut and Alice Wallenberg Foundation (No. 2023.0087), and Chalmers University of Technology, via the Department of Physics and the Areas of Advance Nano and Materials.
The computations were enabled by resources provided by the National Academic Infrastructure for Supercomputing in Sweden (NAISS) at PDC, C3SE, and NSC, partially funded by the Swedish Research Council through grant agreement no. 2022-06725, and the Berzelius resource provided by the Knut and Alice Wallenberg Foundation at NSC.


\begin{thebibliography}{49}%
\makeatletter
\providecommand \@ifxundefined [1]{%
 \@ifx{#1\undefined}
}%
\providecommand \@ifnum [1]{%
 \ifnum #1\expandafter \@firstoftwo
 \else \expandafter \@secondoftwo
 \fi
}%
\providecommand \@ifx [1]{%
 \ifx #1\expandafter \@firstoftwo
 \else \expandafter \@secondoftwo
 \fi
}%
\providecommand \natexlab [1]{#1}%
\providecommand \enquote  [1]{``#1''}%
\providecommand \bibnamefont  [1]{#1}%
\providecommand \bibfnamefont [1]{#1}%
\providecommand \citenamefont [1]{#1}%
\providecommand \href@noop [0]{\@secondoftwo}%
\providecommand \href [0]{\begingroup \@sanitize@url \@href}%
\providecommand \@href[1]{\@@startlink{#1}\@@href}%
\providecommand \@@href[1]{\endgroup#1\@@endlink}%
\providecommand \@sanitize@url [0]{\catcode `\\12\catcode `\$12\catcode `\&12\catcode `\#12\catcode `\^12\catcode `\_12\catcode `\%12\relax}%
\providecommand \@@startlink[1]{}%
\providecommand \@@endlink[0]{}%
\providecommand \url  [0]{\begingroup\@sanitize@url \@url }%
\providecommand \@url [1]{\endgroup\@href {#1}{\urlprefix }}%
\providecommand \urlprefix  [0]{URL }%
\providecommand \Eprint [0]{\href }%
\providecommand \doibase [0]{https://doi.org/}%
\providecommand \selectlanguage [0]{\@gobble}%
\providecommand \bibinfo  [0]{\@secondoftwo}%
\providecommand \bibfield  [0]{\@secondoftwo}%
\providecommand \translation [1]{[#1]}%
\providecommand \BibitemOpen [0]{}%
\providecommand \bibitemStop [0]{}%
\providecommand \bibitemNoStop [0]{.\EOS\space}%
\providecommand \EOS [0]{\spacefactor3000\relax}%
\providecommand \BibitemShut  [1]{\csname bibitem#1\endcsname}%
\let\auto@bib@innerbib\@empty
%</preamble>
\bibitem [{\citenamefont {Disa}\ \emph {et~al.}(2021)\citenamefont {Disa}, \citenamefont {Nova},\ and\ \citenamefont {Cavalleri}}]{Disa2021}%
  \BibitemOpen
  \bibfield  {author} {\bibinfo {author} {\bibfnamefont {A.~S.}\ \bibnamefont {Disa}}, \bibinfo {author} {\bibfnamefont {T.~F.}\ \bibnamefont {Nova}},\ and\ \bibinfo {author} {\bibfnamefont {A.}~\bibnamefont {Cavalleri}},\ }\href {https://doi.org/10.1038/s41567-021-01366-1} {\bibfield  {journal} {\bibinfo  {journal} {Nature Physics}\ }\textbf {\bibinfo {volume} {17}},\ \bibinfo {pages} {1087} (\bibinfo {year} {2021})}\BibitemShut {NoStop}%
\bibitem [{\citenamefont {Salén}\ \emph {et~al.}(2019)\citenamefont {Salén}, \citenamefont {Basini}, \citenamefont {Bonetti}, \citenamefont {Hebling}, \citenamefont {Krasilnikov}, \citenamefont {Nikitin}, \citenamefont {Shamuilov}, \citenamefont {Tibai}, \citenamefont {Zhaunerchyk},\ and\ \citenamefont {Goryashko}}]{Salen2019}%
  \BibitemOpen
  \bibfield  {author} {\bibinfo {author} {\bibfnamefont {P.}~\bibnamefont {Salén}}, \bibinfo {author} {\bibfnamefont {M.}~\bibnamefont {Basini}}, \bibinfo {author} {\bibfnamefont {S.}~\bibnamefont {Bonetti}}, \bibinfo {author} {\bibfnamefont {J.}~\bibnamefont {Hebling}}, \bibinfo {author} {\bibfnamefont {M.}~\bibnamefont {Krasilnikov}}, \bibinfo {author} {\bibfnamefont {A.~Y.}\ \bibnamefont {Nikitin}}, \bibinfo {author} {\bibfnamefont {G.}~\bibnamefont {Shamuilov}}, \bibinfo {author} {\bibfnamefont {Z.}~\bibnamefont {Tibai}}, \bibinfo {author} {\bibfnamefont {V.}~\bibnamefont {Zhaunerchyk}},\ and\ \bibinfo {author} {\bibfnamefont {V.}~\bibnamefont {Goryashko}},\ }\href {https://doi.org/10.1016/j.physrep.2019.09.002} {\bibfield  {journal} {\bibinfo  {journal} {Physics Reports}\ }\textbf {\bibinfo {volume} {836-837}},\ \bibinfo {pages} {1} (\bibinfo {year} {2019})}\BibitemShut {NoStop}%
\bibitem [{\citenamefont {Nicoletti}\ and\ \citenamefont {Cavalleri}(2016)}]{Nicoletti2016}%
  \BibitemOpen
  \bibfield  {author} {\bibinfo {author} {\bibfnamefont {D.}~\bibnamefont {Nicoletti}}\ and\ \bibinfo {author} {\bibfnamefont {A.}~\bibnamefont {Cavalleri}},\ }\href {https://doi.org/10.1364/AOP.8.000401} {\bibfield  {journal} {\bibinfo  {journal} {Advances in Optics and Photonics}\ }\textbf {\bibinfo {volume} {8}},\ \bibinfo {pages} {401} (\bibinfo {year} {2016})}\BibitemShut {NoStop}%
\bibitem [{\citenamefont {Fausti}\ \emph {et~al.}(2011)\citenamefont {Fausti}, \citenamefont {Tobey}, \citenamefont {Dean}, \citenamefont {Kaiser}, \citenamefont {Dienst}, \citenamefont {Hoffmann}, \citenamefont {Pyon}, \citenamefont {Takayama}, \citenamefont {Takagi},\ and\ \citenamefont {Cavalleri}}]{Fausti2011}%
  \BibitemOpen
  \bibfield  {author} {\bibinfo {author} {\bibfnamefont {D.}~\bibnamefont {Fausti}}, \bibinfo {author} {\bibfnamefont {R.~I.}\ \bibnamefont {Tobey}}, \bibinfo {author} {\bibfnamefont {N.}~\bibnamefont {Dean}}, \bibinfo {author} {\bibfnamefont {S.}~\bibnamefont {Kaiser}}, \bibinfo {author} {\bibfnamefont {A.}~\bibnamefont {Dienst}}, \bibinfo {author} {\bibfnamefont {M.~C.}\ \bibnamefont {Hoffmann}}, \bibinfo {author} {\bibfnamefont {S.}~\bibnamefont {Pyon}}, \bibinfo {author} {\bibfnamefont {T.}~\bibnamefont {Takayama}}, \bibinfo {author} {\bibfnamefont {H.}~\bibnamefont {Takagi}},\ and\ \bibinfo {author} {\bibfnamefont {A.}~\bibnamefont {Cavalleri}},\ }\href {https://doi.org/10.1126/science.1197294} {\bibfield  {journal} {\bibinfo  {journal} {Science}\ }\textbf {\bibinfo {volume} {331}},\ \bibinfo {pages} {189} (\bibinfo {year} {2011})}\BibitemShut {NoStop}%
\bibitem [{\citenamefont {Subedi}(2017)}]{Subedi2017}%
  \BibitemOpen
  \bibfield  {author} {\bibinfo {author} {\bibfnamefont {A.}~\bibnamefont {Subedi}},\ }\href {https://doi.org/10.1103/PhysRevB.95.134113} {\bibfield  {journal} {\bibinfo  {journal} {Physical Review B}\ }\textbf {\bibinfo {volume} {95}},\ \bibinfo {pages} {134113} (\bibinfo {year} {2017})}\BibitemShut {NoStop}%
\bibitem [{\citenamefont {Nova}\ \emph {et~al.}(2019)\citenamefont {Nova}, \citenamefont {Disa}, \citenamefont {Fechner},\ and\ \citenamefont {Cavalleri}}]{Nova2019}%
  \BibitemOpen
  \bibfield  {author} {\bibinfo {author} {\bibfnamefont {T.~F.}\ \bibnamefont {Nova}}, \bibinfo {author} {\bibfnamefont {A.~S.}\ \bibnamefont {Disa}}, \bibinfo {author} {\bibfnamefont {M.}~\bibnamefont {Fechner}},\ and\ \bibinfo {author} {\bibfnamefont {A.}~\bibnamefont {Cavalleri}},\ }\href {https://doi.org/10.1126/science.aaw4911} {\bibfield  {journal} {\bibinfo  {journal} {Science}\ }\textbf {\bibinfo {volume} {364}},\ \bibinfo {pages} {1075} (\bibinfo {year} {2019})}\BibitemShut {NoStop}%
\bibitem [{\citenamefont {Buzzi}\ \emph {et~al.}(2021)\citenamefont {Buzzi}, \citenamefont {Nicoletti}, \citenamefont {Fava}, \citenamefont {Jotzu}, \citenamefont {Miyagawa}, \citenamefont {Kanoda}, \citenamefont {Henderson}, \citenamefont {Siegrist}, \citenamefont {Schlueter}, \citenamefont {Nam}, \citenamefont {Ardavan},\ and\ \citenamefont {Cavalleri}}]{Buzzi2021}%
  \BibitemOpen
  \bibfield  {author} {\bibinfo {author} {\bibfnamefont {M.}~\bibnamefont {Buzzi}}, \bibinfo {author} {\bibfnamefont {D.}~\bibnamefont {Nicoletti}}, \bibinfo {author} {\bibfnamefont {S.}~\bibnamefont {Fava}}, \bibinfo {author} {\bibfnamefont {G.}~\bibnamefont {Jotzu}}, \bibinfo {author} {\bibfnamefont {K.}~\bibnamefont {Miyagawa}}, \bibinfo {author} {\bibfnamefont {K.}~\bibnamefont {Kanoda}}, \bibinfo {author} {\bibfnamefont {A.}~\bibnamefont {Henderson}}, \bibinfo {author} {\bibfnamefont {T.}~\bibnamefont {Siegrist}}, \bibinfo {author} {\bibfnamefont {J.~A.}\ \bibnamefont {Schlueter}}, \bibinfo {author} {\bibfnamefont {M.-S.}\ \bibnamefont {Nam}}, \bibinfo {author} {\bibfnamefont {A.}~\bibnamefont {Ardavan}},\ and\ \bibinfo {author} {\bibfnamefont {A.}~\bibnamefont {Cavalleri}},\ }\href {https://doi.org/10.1103/PhysRevLett.127.197002} {\bibfield  {journal} {\bibinfo  {journal} {Physical Review Letters}\ }\textbf {\bibinfo {volume} {127}},\ \bibinfo {pages} {197002} (\bibinfo {year} {2021})}\BibitemShut
  {NoStop}%
\bibitem [{\citenamefont {Budden}\ \emph {et~al.}(2021)\citenamefont {Budden}, \citenamefont {Gebert}, \citenamefont {Buzzi}, \citenamefont {Jotzu}, \citenamefont {Wang}, \citenamefont {Matsuyama}, \citenamefont {Meier}, \citenamefont {Laplace}, \citenamefont {Pontiroli}, \citenamefont {Riccò}, \citenamefont {Schlawin}, \citenamefont {Jaksch},\ and\ \citenamefont {Cavalleri}}]{Budden2021}%
  \BibitemOpen
  \bibfield  {author} {\bibinfo {author} {\bibfnamefont {M.}~\bibnamefont {Budden}}, \bibinfo {author} {\bibfnamefont {T.}~\bibnamefont {Gebert}}, \bibinfo {author} {\bibfnamefont {M.}~\bibnamefont {Buzzi}}, \bibinfo {author} {\bibfnamefont {G.}~\bibnamefont {Jotzu}}, \bibinfo {author} {\bibfnamefont {E.}~\bibnamefont {Wang}}, \bibinfo {author} {\bibfnamefont {T.}~\bibnamefont {Matsuyama}}, \bibinfo {author} {\bibfnamefont {G.}~\bibnamefont {Meier}}, \bibinfo {author} {\bibfnamefont {Y.}~\bibnamefont {Laplace}}, \bibinfo {author} {\bibfnamefont {D.}~\bibnamefont {Pontiroli}}, \bibinfo {author} {\bibfnamefont {M.}~\bibnamefont {Riccò}}, \bibinfo {author} {\bibfnamefont {F.}~\bibnamefont {Schlawin}}, \bibinfo {author} {\bibfnamefont {D.}~\bibnamefont {Jaksch}},\ and\ \bibinfo {author} {\bibfnamefont {A.}~\bibnamefont {Cavalleri}},\ }\href {https://doi.org/10.1038/s41567-020-01148-1} {\bibfield  {journal} {\bibinfo  {journal} {Nature Physics}\ }\textbf {\bibinfo {volume} {17}},\ \bibinfo {pages} {611} (\bibinfo
  {year} {2021})}\BibitemShut {NoStop}%
\bibitem [{\citenamefont {Latini}\ \emph {et~al.}(2021)\citenamefont {Latini}, \citenamefont {Shin}, \citenamefont {Sato}, \citenamefont {Schäfer}, \citenamefont {Giovannini}, \citenamefont {Hübener},\ and\ \citenamefont {Rubio}}]{Latini2021}%
  \BibitemOpen
  \bibfield  {author} {\bibinfo {author} {\bibfnamefont {S.}~\bibnamefont {Latini}}, \bibinfo {author} {\bibfnamefont {D.}~\bibnamefont {Shin}}, \bibinfo {author} {\bibfnamefont {S.~A.}\ \bibnamefont {Sato}}, \bibinfo {author} {\bibfnamefont {C.}~\bibnamefont {Schäfer}}, \bibinfo {author} {\bibfnamefont {U.~D.}\ \bibnamefont {Giovannini}}, \bibinfo {author} {\bibfnamefont {H.}~\bibnamefont {Hübener}},\ and\ \bibinfo {author} {\bibfnamefont {A.}~\bibnamefont {Rubio}},\ }\bibfield  {journal} {\bibinfo  {journal} {Proceedings of the National Academy of Sciences}\ }\textbf {\bibinfo {volume} {118}},\ \href {https://doi.org/10.1073/pnas.2105618118} {10.1073/pnas.2105618118} (\bibinfo {year} {2021})\BibitemShut {NoStop}%
\bibitem [{\citenamefont {Cheng}\ \emph {et~al.}(2023)\citenamefont {Cheng}, \citenamefont {Kramer}, \citenamefont {Shen},\ and\ \citenamefont {Hoffmann}}]{Cheng2023}%
  \BibitemOpen
  \bibfield  {author} {\bibinfo {author} {\bibfnamefont {B.}~\bibnamefont {Cheng}}, \bibinfo {author} {\bibfnamefont {P.~L.}\ \bibnamefont {Kramer}}, \bibinfo {author} {\bibfnamefont {Z.-X.}\ \bibnamefont {Shen}},\ and\ \bibinfo {author} {\bibfnamefont {M.~C.}\ \bibnamefont {Hoffmann}},\ }\href {https://doi.org/10.1103/PhysRevLett.130.126902} {\bibfield  {journal} {\bibinfo  {journal} {Physical Review Letters}\ }\textbf {\bibinfo {volume} {130}},\ \bibinfo {pages} {126902} (\bibinfo {year} {2023})}\BibitemShut {NoStop}%
\bibitem [{\citenamefont {Juraschek}\ \emph {et~al.}(2017)\citenamefont {Juraschek}, \citenamefont {Fechner}, \citenamefont {Balatsky},\ and\ \citenamefont {Spaldin}}]{Juraschek2017}%
  \BibitemOpen
  \bibfield  {author} {\bibinfo {author} {\bibfnamefont {D.~M.}\ \bibnamefont {Juraschek}}, \bibinfo {author} {\bibfnamefont {M.}~\bibnamefont {Fechner}}, \bibinfo {author} {\bibfnamefont {A.~V.}\ \bibnamefont {Balatsky}},\ and\ \bibinfo {author} {\bibfnamefont {N.~A.}\ \bibnamefont {Spaldin}},\ }\href {https://doi.org/10.1103/PhysRevMaterials.1.014401} {\bibfield  {journal} {\bibinfo  {journal} {Physical Review Materials}\ }\textbf {\bibinfo {volume} {1}},\ \bibinfo {pages} {014401} (\bibinfo {year} {2017})}\BibitemShut {NoStop}%
\bibitem [{\citenamefont {Juraschek}\ and\ \citenamefont {Spaldin}(2019)}]{Juraschek2019}%
  \BibitemOpen
  \bibfield  {author} {\bibinfo {author} {\bibfnamefont {D.~M.}\ \bibnamefont {Juraschek}}\ and\ \bibinfo {author} {\bibfnamefont {N.~A.}\ \bibnamefont {Spaldin}},\ }\href {https://doi.org/10.1103/PhysRevMaterials.3.064405} {\bibfield  {journal} {\bibinfo  {journal} {Physical Review Materials}\ }\textbf {\bibinfo {volume} {3}},\ \bibinfo {pages} {064405} (\bibinfo {year} {2019})}\BibitemShut {NoStop}%
\bibitem [{\citenamefont {Cheng}\ \emph {et~al.}(2020)\citenamefont {Cheng}, \citenamefont {Schumann}, \citenamefont {Wang}, \citenamefont {Zhang}, \citenamefont {Barbalas}, \citenamefont {Stemmer},\ and\ \citenamefont {Armitage}}]{Cheng2020}%
  \BibitemOpen
  \bibfield  {author} {\bibinfo {author} {\bibfnamefont {B.}~\bibnamefont {Cheng}}, \bibinfo {author} {\bibfnamefont {T.}~\bibnamefont {Schumann}}, \bibinfo {author} {\bibfnamefont {Y.}~\bibnamefont {Wang}}, \bibinfo {author} {\bibfnamefont {X.}~\bibnamefont {Zhang}}, \bibinfo {author} {\bibfnamefont {D.}~\bibnamefont {Barbalas}}, \bibinfo {author} {\bibfnamefont {S.}~\bibnamefont {Stemmer}},\ and\ \bibinfo {author} {\bibfnamefont {N.~P.}\ \bibnamefont {Armitage}},\ }\href {https://doi.org/10.1021/acs.nanolett.0c01983} {\bibfield  {journal} {\bibinfo  {journal} {Nano Letters}\ }\textbf {\bibinfo {volume} {20}},\ \bibinfo {pages} {5991} (\bibinfo {year} {2020})}\BibitemShut {NoStop}%
\bibitem [{\citenamefont {Geilhufe}\ \emph {et~al.}(2021)\citenamefont {Geilhufe}, \citenamefont {Juričić}, \citenamefont {Bonetti}, \citenamefont {Zhu},\ and\ \citenamefont {Balatsky}}]{Geilhufe2021}%
  \BibitemOpen
  \bibfield  {author} {\bibinfo {author} {\bibfnamefont {R.~M.}\ \bibnamefont {Geilhufe}}, \bibinfo {author} {\bibfnamefont {V.}~\bibnamefont {Juričić}}, \bibinfo {author} {\bibfnamefont {S.}~\bibnamefont {Bonetti}}, \bibinfo {author} {\bibfnamefont {J.-X.}\ \bibnamefont {Zhu}},\ and\ \bibinfo {author} {\bibfnamefont {A.~V.}\ \bibnamefont {Balatsky}},\ }\href {https://doi.org/10.1103/PhysRevResearch.3.L022011} {\bibfield  {journal} {\bibinfo  {journal} {Physical Review Research}\ }\textbf {\bibinfo {volume} {3}},\ \bibinfo {pages} {L022011} (\bibinfo {year} {2021})}\BibitemShut {NoStop}%
\bibitem [{\citenamefont {Ren}\ \emph {et~al.}(2021)\citenamefont {Ren}, \citenamefont {Xiao}, \citenamefont {Saparov},\ and\ \citenamefont {Niu}}]{Ren2021}%
  \BibitemOpen
  \bibfield  {author} {\bibinfo {author} {\bibfnamefont {Y.}~\bibnamefont {Ren}}, \bibinfo {author} {\bibfnamefont {C.}~\bibnamefont {Xiao}}, \bibinfo {author} {\bibfnamefont {D.}~\bibnamefont {Saparov}},\ and\ \bibinfo {author} {\bibfnamefont {Q.}~\bibnamefont {Niu}},\ }\href {https://doi.org/10.1103/PhysRevLett.127.186403} {\bibfield  {journal} {\bibinfo  {journal} {Physical Review Letters}\ }\textbf {\bibinfo {volume} {127}},\ \bibinfo {pages} {186403} (\bibinfo {year} {2021})}\BibitemShut {NoStop}%
\bibitem [{\citenamefont {Juraschek}\ \emph {et~al.}(2022)\citenamefont {Juraschek}, \citenamefont {Neuman},\ and\ \citenamefont {Narang}}]{Juraschek2022}%
  \BibitemOpen
  \bibfield  {author} {\bibinfo {author} {\bibfnamefont {D.~M.}\ \bibnamefont {Juraschek}}, \bibinfo {author} {\bibfnamefont {T.}~\bibnamefont {Neuman}},\ and\ \bibinfo {author} {\bibfnamefont {P.}~\bibnamefont {Narang}},\ }\href {https://doi.org/10.1103/PhysRevResearch.4.013129} {\bibfield  {journal} {\bibinfo  {journal} {Physical Review Research}\ }\textbf {\bibinfo {volume} {4}},\ \bibinfo {pages} {013129} (\bibinfo {year} {2022})}\BibitemShut {NoStop}%
\bibitem [{\citenamefont {Baydin}\ \emph {et~al.}(2022)\citenamefont {Baydin}, \citenamefont {Hernandez}, \citenamefont {Rodriguez-Vega}, \citenamefont {Okazaki}, \citenamefont {Tay}, \citenamefont {Noe}, \citenamefont {Katayama}, \citenamefont {Takeda}, \citenamefont {Nojiri}, \citenamefont {Rappl}, \citenamefont {Abramof}, \citenamefont {Fiete},\ and\ \citenamefont {Kono}}]{Baydin2022}%
  \BibitemOpen
  \bibfield  {author} {\bibinfo {author} {\bibfnamefont {A.}~\bibnamefont {Baydin}}, \bibinfo {author} {\bibfnamefont {F.~G.~G.}\ \bibnamefont {Hernandez}}, \bibinfo {author} {\bibfnamefont {M.}~\bibnamefont {Rodriguez-Vega}}, \bibinfo {author} {\bibfnamefont {A.~K.}\ \bibnamefont {Okazaki}}, \bibinfo {author} {\bibfnamefont {F.}~\bibnamefont {Tay}}, \bibinfo {author} {\bibfnamefont {G.~T.}\ \bibnamefont {Noe}}, \bibinfo {author} {\bibfnamefont {I.}~\bibnamefont {Katayama}}, \bibinfo {author} {\bibfnamefont {J.}~\bibnamefont {Takeda}}, \bibinfo {author} {\bibfnamefont {H.}~\bibnamefont {Nojiri}}, \bibinfo {author} {\bibfnamefont {P.~H.~O.}\ \bibnamefont {Rappl}}, \bibinfo {author} {\bibfnamefont {E.}~\bibnamefont {Abramof}}, \bibinfo {author} {\bibfnamefont {G.~A.}\ \bibnamefont {Fiete}},\ and\ \bibinfo {author} {\bibfnamefont {J.}~\bibnamefont {Kono}},\ }\href {https://doi.org/10.1103/PhysRevLett.128.075901} {\bibfield  {journal} {\bibinfo  {journal} {Physical Review Letters}\ }\textbf {\bibinfo {volume}
  {128}},\ \bibinfo {pages} {075901} (\bibinfo {year} {2022})}\BibitemShut {NoStop}%
\bibitem [{\citenamefont {Basini}\ \emph {et~al.}(2024)\citenamefont {Basini}, \citenamefont {Pancaldi}, \citenamefont {Wehinger}, \citenamefont {Udina}, \citenamefont {Unikandanunni}, \citenamefont {Tadano}, \citenamefont {Hoffmann}, \citenamefont {Balatsky},\ and\ \citenamefont {Bonetti}}]{Basini2022}%
  \BibitemOpen
  \bibfield  {author} {\bibinfo {author} {\bibfnamefont {M.}~\bibnamefont {Basini}}, \bibinfo {author} {\bibfnamefont {M.}~\bibnamefont {Pancaldi}}, \bibinfo {author} {\bibfnamefont {B.}~\bibnamefont {Wehinger}}, \bibinfo {author} {\bibfnamefont {M.}~\bibnamefont {Udina}}, \bibinfo {author} {\bibfnamefont {V.}~\bibnamefont {Unikandanunni}}, \bibinfo {author} {\bibfnamefont {T.}~\bibnamefont {Tadano}}, \bibinfo {author} {\bibfnamefont {M.~C.}\ \bibnamefont {Hoffmann}}, \bibinfo {author} {\bibfnamefont {A.~V.}\ \bibnamefont {Balatsky}},\ and\ \bibinfo {author} {\bibfnamefont {S.}~\bibnamefont {Bonetti}},\ }\href {https://doi.org/10.1038/s41586-024-07175-9} {\bibfield  {journal} {\bibinfo  {journal} {Nature}\ }\textbf {\bibinfo {volume} {628}},\ \bibinfo {pages} {534} (\bibinfo {year} {2024})}\BibitemShut {NoStop}%
\bibitem [{\citenamefont {Geilhufe}\ and\ \citenamefont {Hergert}(2023)}]{Geilhufe2023}%
  \BibitemOpen
  \bibfield  {author} {\bibinfo {author} {\bibfnamefont {R.~M.}\ \bibnamefont {Geilhufe}}\ and\ \bibinfo {author} {\bibfnamefont {W.}~\bibnamefont {Hergert}},\ }\href {https://doi.org/10.1103/PhysRevB.107.L020406} {\bibfield  {journal} {\bibinfo  {journal} {Physical Review B}\ }\textbf {\bibinfo {volume} {107}},\ \bibinfo {pages} {L020406} (\bibinfo {year} {2023})}\BibitemShut {NoStop}%
\bibitem [{\citenamefont {Luo}\ \emph {et~al.}(2023)\citenamefont {Luo}, \citenamefont {Lin}, \citenamefont {Zhang}, \citenamefont {Chen}, \citenamefont {Blackert}, \citenamefont {Xu}, \citenamefont {Yakobson},\ and\ \citenamefont {Zhu}}]{Luo2023}%
  \BibitemOpen
  \bibfield  {author} {\bibinfo {author} {\bibfnamefont {J.}~\bibnamefont {Luo}}, \bibinfo {author} {\bibfnamefont {T.}~\bibnamefont {Lin}}, \bibinfo {author} {\bibfnamefont {J.}~\bibnamefont {Zhang}}, \bibinfo {author} {\bibfnamefont {X.}~\bibnamefont {Chen}}, \bibinfo {author} {\bibfnamefont {E.~R.}\ \bibnamefont {Blackert}}, \bibinfo {author} {\bibfnamefont {R.}~\bibnamefont {Xu}}, \bibinfo {author} {\bibfnamefont {B.~I.}\ \bibnamefont {Yakobson}},\ and\ \bibinfo {author} {\bibfnamefont {H.}~\bibnamefont {Zhu}},\ }\href {https://doi.org/10.1126/science.adi9601} {\bibfield  {journal} {\bibinfo  {journal} {Science}\ }\textbf {\bibinfo {volume} {382}},\ \bibinfo {pages} {698} (\bibinfo {year} {2023})}\BibitemShut {NoStop}%
\bibitem [{\citenamefont {Hernandez}\ \emph {et~al.}(2023)\citenamefont {Hernandez}, \citenamefont {Baydin}, \citenamefont {Chaudhary}, \citenamefont {Tay}, \citenamefont {Katayama}, \citenamefont {Takeda}, \citenamefont {Nojiri}, \citenamefont {Okazaki}, \citenamefont {Rappl}, \citenamefont {Abramof}, \citenamefont {Rodriguez-Vega}, \citenamefont {Fiete},\ and\ \citenamefont {Kono}}]{Hernandez2023}%
  \BibitemOpen
  \bibfield  {author} {\bibinfo {author} {\bibfnamefont {F.~G.~G.}\ \bibnamefont {Hernandez}}, \bibinfo {author} {\bibfnamefont {A.}~\bibnamefont {Baydin}}, \bibinfo {author} {\bibfnamefont {S.}~\bibnamefont {Chaudhary}}, \bibinfo {author} {\bibfnamefont {F.}~\bibnamefont {Tay}}, \bibinfo {author} {\bibfnamefont {I.}~\bibnamefont {Katayama}}, \bibinfo {author} {\bibfnamefont {J.}~\bibnamefont {Takeda}}, \bibinfo {author} {\bibfnamefont {H.}~\bibnamefont {Nojiri}}, \bibinfo {author} {\bibfnamefont {A.~K.}\ \bibnamefont {Okazaki}}, \bibinfo {author} {\bibfnamefont {P.~H.~O.}\ \bibnamefont {Rappl}}, \bibinfo {author} {\bibfnamefont {E.}~\bibnamefont {Abramof}}, \bibinfo {author} {\bibfnamefont {M.}~\bibnamefont {Rodriguez-Vega}}, \bibinfo {author} {\bibfnamefont {G.~A.}\ \bibnamefont {Fiete}},\ and\ \bibinfo {author} {\bibfnamefont {J.}~\bibnamefont {Kono}},\ }\href {https://doi.org/10.1126/sciadv.adj4074} {\bibfield  {journal} {\bibinfo  {journal} {Science Advances}\ }\textbf {\bibinfo {volume} {9}},\ \bibinfo
  {pages} {eadj407} (\bibinfo {year} {2023})}\BibitemShut {NoStop}%
\bibitem [{\citenamefont {Shabala}\ and\ \citenamefont {Geilhufe}(2024)}]{Shabala2024}%
  \BibitemOpen
  \bibfield  {author} {\bibinfo {author} {\bibfnamefont {N.}~\bibnamefont {Shabala}}\ and\ \bibinfo {author} {\bibfnamefont {R.~M.}\ \bibnamefont {Geilhufe}},\ }\href {https://doi.org/10.1103/PhysRevLett.133.266702} {\bibfield  {journal} {\bibinfo  {journal} {Physical Review Letters}\ }\textbf {\bibinfo {volume} {133}},\ \bibinfo {pages} {266702} (\bibinfo {year} {2024})}\BibitemShut {NoStop}%
\bibitem [{\citenamefont {Shabala}\ \emph {et~al.}(2025)\citenamefont {Shabala}, \citenamefont {Tietjen},\ and\ \citenamefont {Geilhufe}}]{Shabala2025}%
  \BibitemOpen
  \bibfield  {author} {\bibinfo {author} {\bibfnamefont {N.}~\bibnamefont {Shabala}}, \bibinfo {author} {\bibfnamefont {F.}~\bibnamefont {Tietjen}},\ and\ \bibinfo {author} {\bibfnamefont {R.~M.}\ \bibnamefont {Geilhufe}},\ }\href {https://doi.org/10.48550/arXiv.2511.03329} {\bibfield  {journal} {\bibinfo  {journal} {arXiv:2511.03329}\ } (\bibinfo {year} {2025})}\BibitemShut {NoStop}%
\bibitem [{\citenamefont {Disa}\ \emph {et~al.}(2020)\citenamefont {Disa}, \citenamefont {Fechner}, \citenamefont {Nova}, \citenamefont {Liu}, \citenamefont {Först}, \citenamefont {Prabhakaran}, \citenamefont {Radaelli},\ and\ \citenamefont {Cavalleri}}]{Disa2020}%
  \BibitemOpen
  \bibfield  {author} {\bibinfo {author} {\bibfnamefont {A.~S.}\ \bibnamefont {Disa}}, \bibinfo {author} {\bibfnamefont {M.}~\bibnamefont {Fechner}}, \bibinfo {author} {\bibfnamefont {T.~F.}\ \bibnamefont {Nova}}, \bibinfo {author} {\bibfnamefont {B.}~\bibnamefont {Liu}}, \bibinfo {author} {\bibfnamefont {M.}~\bibnamefont {Först}}, \bibinfo {author} {\bibfnamefont {D.}~\bibnamefont {Prabhakaran}}, \bibinfo {author} {\bibfnamefont {P.~G.}\ \bibnamefont {Radaelli}},\ and\ \bibinfo {author} {\bibfnamefont {A.}~\bibnamefont {Cavalleri}},\ }\href {https://doi.org/10.1038/s41567-020-0936-3} {\bibfield  {journal} {\bibinfo  {journal} {Nature Physics}\ }\textbf {\bibinfo {volume} {16}},\ \bibinfo {pages} {937} (\bibinfo {year} {2020})}\BibitemShut {NoStop}%
\bibitem [{\citenamefont {Davies}\ \emph {et~al.}(2024)\citenamefont {Davies}, \citenamefont {Fennema}, \citenamefont {Tsukamoto}, \citenamefont {Razdolski}, \citenamefont {Kimel},\ and\ \citenamefont {Kirilyuk}}]{Davies2023}%
  \BibitemOpen
  \bibfield  {author} {\bibinfo {author} {\bibfnamefont {C.~S.}\ \bibnamefont {Davies}}, \bibinfo {author} {\bibfnamefont {F.~G.~N.}\ \bibnamefont {Fennema}}, \bibinfo {author} {\bibfnamefont {A.}~\bibnamefont {Tsukamoto}}, \bibinfo {author} {\bibfnamefont {I.}~\bibnamefont {Razdolski}}, \bibinfo {author} {\bibfnamefont {A.~V.}\ \bibnamefont {Kimel}},\ and\ \bibinfo {author} {\bibfnamefont {A.}~\bibnamefont {Kirilyuk}},\ }\href {https://doi.org/10.1038/s41586-024-07200-x} {\bibfield  {journal} {\bibinfo  {journal} {Nature}\ }\textbf {\bibinfo {volume} {628}},\ \bibinfo {pages} {540} (\bibinfo {year} {2024})}\BibitemShut {NoStop}%
\bibitem [{\citenamefont {Caprini}\ \emph {et~al.}(2024)\citenamefont {Caprini}, \citenamefont {Löwen},\ and\ \citenamefont {Geilhufe}}]{Caprini2024}%
  \BibitemOpen
  \bibfield  {author} {\bibinfo {author} {\bibfnamefont {L.}~\bibnamefont {Caprini}}, \bibinfo {author} {\bibfnamefont {H.}~\bibnamefont {Löwen}},\ and\ \bibinfo {author} {\bibfnamefont {R.~M.}\ \bibnamefont {Geilhufe}},\ }\href {https://doi.org/10.1038/s41467-023-44277-w} {\bibfield  {journal} {\bibinfo  {journal} {Nature Communications}\ }\textbf {\bibinfo {volume} {15}},\ \bibinfo {pages} {94} (\bibinfo {year} {2024})}\BibitemShut {NoStop}%
\bibitem [{\citenamefont {Qiao}\ and\ \citenamefont {Geilhufe}(2025)}]{Qiao2025}%
  \BibitemOpen
  \bibfield  {author} {\bibinfo {author} {\bibfnamefont {Y.}~\bibnamefont {Qiao}}\ and\ \bibinfo {author} {\bibfnamefont {R.~M.}\ \bibnamefont {Geilhufe}},\ }\href {https://doi.org/10.48550/arXiv.2512.17669} {\bibfield  {journal} {\bibinfo  {journal} {arXiv:2512.17669}\ } (\bibinfo {year} {2025})}\BibitemShut {NoStop}%
\bibitem [{\citenamefont {Tietjen}\ and\ \citenamefont {Geilhufe}(2025)}]{Tietjen2025}%
  \BibitemOpen
  \bibfield  {author} {\bibinfo {author} {\bibfnamefont {F.}~\bibnamefont {Tietjen}}\ and\ \bibinfo {author} {\bibfnamefont {R.~M.}\ \bibnamefont {Geilhufe}},\ }\href {https://doi.org/10.1093/pnasnexus/pgaf055} {\bibfield  {journal} {\bibinfo  {journal} {PNAS Nexus}\ }\textbf {\bibinfo {volume} {4}} (\bibinfo {year} {2025})}\BibitemShut {NoStop}%
\bibitem [{\citenamefont {Kozina}\ \emph {et~al.}(2019)\citenamefont {Kozina}, \citenamefont {Fechner}, \citenamefont {Marsik}, \citenamefont {van Driel}, \citenamefont {Glownia}, \citenamefont {Bernhard}, \citenamefont {Radovic}, \citenamefont {Zhu}, \citenamefont {Bonetti}, \citenamefont {Staub},\ and\ \citenamefont {Hoffmann}}]{Kozina2019}%
  \BibitemOpen
  \bibfield  {author} {\bibinfo {author} {\bibfnamefont {M.}~\bibnamefont {Kozina}}, \bibinfo {author} {\bibfnamefont {M.}~\bibnamefont {Fechner}}, \bibinfo {author} {\bibfnamefont {P.}~\bibnamefont {Marsik}}, \bibinfo {author} {\bibfnamefont {T.}~\bibnamefont {van Driel}}, \bibinfo {author} {\bibfnamefont {J.~M.}\ \bibnamefont {Glownia}}, \bibinfo {author} {\bibfnamefont {C.}~\bibnamefont {Bernhard}}, \bibinfo {author} {\bibfnamefont {M.}~\bibnamefont {Radovic}}, \bibinfo {author} {\bibfnamefont {D.}~\bibnamefont {Zhu}}, \bibinfo {author} {\bibfnamefont {S.}~\bibnamefont {Bonetti}}, \bibinfo {author} {\bibfnamefont {U.}~\bibnamefont {Staub}},\ and\ \bibinfo {author} {\bibfnamefont {M.~C.}\ \bibnamefont {Hoffmann}},\ }\href {https://doi.org/10.1038/s41567-018-0408-1} {\bibfield  {journal} {\bibinfo  {journal} {Nature Physics}\ }\textbf {\bibinfo {volume} {15}},\ \bibinfo {pages} {387} (\bibinfo {year} {2019})}\BibitemShut {NoStop}%
\bibitem [{\citenamefont {Fleury}\ \emph {et~al.}(1968)\citenamefont {Fleury}, \citenamefont {Scott},\ and\ \citenamefont {Worlock}}]{Fleury1968}%
  \BibitemOpen
  \bibfield  {author} {\bibinfo {author} {\bibfnamefont {P.~A.}\ \bibnamefont {Fleury}}, \bibinfo {author} {\bibfnamefont {J.~F.}\ \bibnamefont {Scott}},\ and\ \bibinfo {author} {\bibfnamefont {J.~M.}\ \bibnamefont {Worlock}},\ }\href {https://doi.org/10.1103/physrevlett.21.16} {\bibfield  {journal} {\bibinfo  {journal} {Physical Review Letters}\ }\textbf {\bibinfo {volume} {21}},\ \bibinfo {pages} {16} (\bibinfo {year} {1968})}\BibitemShut {NoStop}%
\bibitem [{\citenamefont {Cowley}(1996)}]{Cowley1996}%
  \BibitemOpen
  \bibfield  {author} {\bibinfo {author} {\bibfnamefont {R.~A.}\ \bibnamefont {Cowley}},\ }\href {https://doi.org/10.1098/rsta.1996.0130} {\bibfield  {journal} {\bibinfo  {journal} {Philosophical Transactions of the Royal Society of London. Series A: Mathematical, Physical and Engineering Sciences}\ }\textbf {\bibinfo {volume} {354}},\ \bibinfo {pages} {2799} (\bibinfo {year} {1996})}\BibitemShut {NoStop}%
\bibitem [{\citenamefont {Holt}\ \emph {et~al.}(2007)\citenamefont {Holt}, \citenamefont {Sutton}, \citenamefont {Zschack}, \citenamefont {Hong},\ and\ \citenamefont {Chiang}}]{Holt2007}%
  \BibitemOpen
  \bibfield  {author} {\bibinfo {author} {\bibfnamefont {M.}~\bibnamefont {Holt}}, \bibinfo {author} {\bibfnamefont {M.}~\bibnamefont {Sutton}}, \bibinfo {author} {\bibfnamefont {P.}~\bibnamefont {Zschack}}, \bibinfo {author} {\bibfnamefont {H.}~\bibnamefont {Hong}},\ and\ \bibinfo {author} {\bibfnamefont {T.-C.}\ \bibnamefont {Chiang}},\ }\href {https://doi.org/10.1103/physrevlett.98.065501} {\bibfield  {journal} {\bibinfo  {journal} {Physical Review Letters}\ }\textbf {\bibinfo {volume} {98}},\ \bibinfo {pages} {065501} (\bibinfo {year} {2007})}\BibitemShut {NoStop}%
\bibitem [{\citenamefont {M{\"u}ller}\ and\ \citenamefont {Burkard}(1979)}]{muller1979srti}%
  \BibitemOpen
  \bibfield  {author} {\bibinfo {author} {\bibfnamefont {K.~A.}\ \bibnamefont {M{\"u}ller}}\ and\ \bibinfo {author} {\bibfnamefont {H.}~\bibnamefont {Burkard}},\ }\href {https://doi.org/10.1103/PhysRevB.19.3593} {\bibfield  {journal} {\bibinfo  {journal} {Physical Review B}\ }\textbf {\bibinfo {volume} {19}},\ \bibinfo {pages} {3593} (\bibinfo {year} {1979})}\BibitemShut {NoStop}%
\bibitem [{\citenamefont {Verdi}\ \emph {et~al.}(2023)\citenamefont {Verdi}, \citenamefont {Ranalli}, \citenamefont {Franchini},\ and\ \citenamefont {Kresse}}]{VerRanFra23}%
  \BibitemOpen
  \bibfield  {author} {\bibinfo {author} {\bibfnamefont {C.}~\bibnamefont {Verdi}}, \bibinfo {author} {\bibfnamefont {L.}~\bibnamefont {Ranalli}}, \bibinfo {author} {\bibfnamefont {C.}~\bibnamefont {Franchini}},\ and\ \bibinfo {author} {\bibfnamefont {G.}~\bibnamefont {Kresse}},\ }\href {https://doi.org/10.1103/PhysRevMaterials.7.L030801} {\bibfield  {journal} {\bibinfo  {journal} {Physical Review Materials}\ }\textbf {\bibinfo {volume} {7}},\ \bibinfo {pages} {L030801} (\bibinfo {year} {2023})}\BibitemShut {NoStop}%
\bibitem [{\citenamefont {Fan}\ \emph {et~al.}(2022)\citenamefont {Fan}, \citenamefont {Wang}, \citenamefont {Ying}, \citenamefont {Song}, \citenamefont {Wang}, \citenamefont {Wang}, \citenamefont {Zeng}, \citenamefont {Xu}, \citenamefont {Lindgren}, \citenamefont {Rahm}, \citenamefont {Gabourie}, \citenamefont {Liu}, \citenamefont {Dong}, \citenamefont {Wu}, \citenamefont {Chen}, \citenamefont {Zhong}, \citenamefont {Sun}, \citenamefont {Erhart}, \citenamefont {Su},\ and\ \citenamefont {{Ala-Nissila}}}]{FanWanYin22}%
  \BibitemOpen
  \bibfield  {author} {\bibinfo {author} {\bibfnamefont {Z.}~\bibnamefont {Fan}}, \bibinfo {author} {\bibfnamefont {Y.}~\bibnamefont {Wang}}, \bibinfo {author} {\bibfnamefont {P.}~\bibnamefont {Ying}}, \bibinfo {author} {\bibfnamefont {K.}~\bibnamefont {Song}}, \bibinfo {author} {\bibfnamefont {J.}~\bibnamefont {Wang}}, \bibinfo {author} {\bibfnamefont {Y.}~\bibnamefont {Wang}}, \bibinfo {author} {\bibfnamefont {Z.}~\bibnamefont {Zeng}}, \bibinfo {author} {\bibfnamefont {K.}~\bibnamefont {Xu}}, \bibinfo {author} {\bibfnamefont {E.}~\bibnamefont {Lindgren}}, \bibinfo {author} {\bibfnamefont {J.~M.}\ \bibnamefont {Rahm}}, \bibinfo {author} {\bibfnamefont {A.~J.}\ \bibnamefont {Gabourie}}, \bibinfo {author} {\bibfnamefont {J.}~\bibnamefont {Liu}}, \bibinfo {author} {\bibfnamefont {H.}~\bibnamefont {Dong}}, \bibinfo {author} {\bibfnamefont {J.}~\bibnamefont {Wu}}, \bibinfo {author} {\bibfnamefont {Y.}~\bibnamefont {Chen}}, \bibinfo {author} {\bibfnamefont {Z.}~\bibnamefont {Zhong}}, \bibinfo {author}
  {\bibfnamefont {J.}~\bibnamefont {Sun}}, \bibinfo {author} {\bibfnamefont {P.}~\bibnamefont {Erhart}}, \bibinfo {author} {\bibfnamefont {Y.}~\bibnamefont {Su}},\ and\ \bibinfo {author} {\bibfnamefont {T.}~\bibnamefont {{Ala-Nissila}}},\ }\href {https://doi.org/10.1063/5.0106617} {\bibfield  {journal} {\bibinfo  {journal} {The Journal of Chemical Physics}\ }\textbf {\bibinfo {volume} {157}},\ \bibinfo {pages} {114801} (\bibinfo {year} {2022})}\BibitemShut {NoStop}%
\bibitem [{\citenamefont {Xu}\ \emph {et~al.}(2025)\citenamefont {Xu}, \citenamefont {Bu}, \citenamefont {Pan}, \citenamefont {Lindgren}, \citenamefont {Wu}, \citenamefont {Wang}, \citenamefont {Liu}, \citenamefont {Song}, \citenamefont {Xu}, \citenamefont {Li}, \citenamefont {Hainer}, \citenamefont {Svensson}, \citenamefont {Wiktor}, \citenamefont {Zhao}, \citenamefont {Huang}, \citenamefont {Qian}, \citenamefont {Zhang}, \citenamefont {Zeng}, \citenamefont {Zhang}, \citenamefont {Tang}, \citenamefont {Xiao}, \citenamefont {Yan}, \citenamefont {Shi}, \citenamefont {Liang}, \citenamefont {Wang}, \citenamefont {Liang}, \citenamefont {Cao}, \citenamefont {Wang}, \citenamefont {Ying}, \citenamefont {Xu}, \citenamefont {Chen}, \citenamefont {Zhang}, \citenamefont {Chen}, \citenamefont {Wu}, \citenamefont {Jiang}, \citenamefont {Berger}, \citenamefont {Li}, \citenamefont {Chen}, \citenamefont {Gabourie}, \citenamefont {Dong}, \citenamefont {Xiong}, \citenamefont {Wei}, \citenamefont {Chen}, \citenamefont {Xu},
  \citenamefont {Ding}, \citenamefont {Sun}, \citenamefont {Ala-Nissila}, \citenamefont {Harju}, \citenamefont {Zheng}, \citenamefont {Guan}, \citenamefont {Erhart}, \citenamefont {Sun}, \citenamefont {Ouyang}, \citenamefont {Su},\ and\ \citenamefont {Fan}}]{xu2025mega}%
  \BibitemOpen
  \bibfield  {author} {\bibinfo {author} {\bibfnamefont {K.}~\bibnamefont {Xu}}, \bibinfo {author} {\bibfnamefont {H.}~\bibnamefont {Bu}}, \bibinfo {author} {\bibfnamefont {S.}~\bibnamefont {Pan}}, \bibinfo {author} {\bibfnamefont {E.}~\bibnamefont {Lindgren}}, \bibinfo {author} {\bibfnamefont {Y.}~\bibnamefont {Wu}}, \bibinfo {author} {\bibfnamefont {Y.}~\bibnamefont {Wang}}, \bibinfo {author} {\bibfnamefont {J.}~\bibnamefont {Liu}}, \bibinfo {author} {\bibfnamefont {K.}~\bibnamefont {Song}}, \bibinfo {author} {\bibfnamefont {B.}~\bibnamefont {Xu}}, \bibinfo {author} {\bibfnamefont {Y.}~\bibnamefont {Li}}, \bibinfo {author} {\bibfnamefont {T.}~\bibnamefont {Hainer}}, \bibinfo {author} {\bibfnamefont {L.}~\bibnamefont {Svensson}}, \bibinfo {author} {\bibfnamefont {J.}~\bibnamefont {Wiktor}}, \bibinfo {author} {\bibfnamefont {R.}~\bibnamefont {Zhao}}, \bibinfo {author} {\bibfnamefont {H.}~\bibnamefont {Huang}}, \bibinfo {author} {\bibfnamefont {C.}~\bibnamefont {Qian}}, \bibinfo {author} {\bibfnamefont
  {S.}~\bibnamefont {Zhang}}, \bibinfo {author} {\bibfnamefont {Z.}~\bibnamefont {Zeng}}, \bibinfo {author} {\bibfnamefont {B.}~\bibnamefont {Zhang}}, \bibinfo {author} {\bibfnamefont {B.}~\bibnamefont {Tang}}, \bibinfo {author} {\bibfnamefont {Y.}~\bibnamefont {Xiao}}, \bibinfo {author} {\bibfnamefont {Z.}~\bibnamefont {Yan}}, \bibinfo {author} {\bibfnamefont {J.}~\bibnamefont {Shi}}, \bibinfo {author} {\bibfnamefont {Z.}~\bibnamefont {Liang}}, \bibinfo {author} {\bibfnamefont {J.}~\bibnamefont {Wang}}, \bibinfo {author} {\bibfnamefont {T.}~\bibnamefont {Liang}}, \bibinfo {author} {\bibfnamefont {S.}~\bibnamefont {Cao}}, \bibinfo {author} {\bibfnamefont {Y.}~\bibnamefont {Wang}}, \bibinfo {author} {\bibfnamefont {P.}~\bibnamefont {Ying}}, \bibinfo {author} {\bibfnamefont {N.}~\bibnamefont {Xu}}, \bibinfo {author} {\bibfnamefont {C.}~\bibnamefont {Chen}}, \bibinfo {author} {\bibfnamefont {Y.}~\bibnamefont {Zhang}}, \bibinfo {author} {\bibfnamefont {Z.}~\bibnamefont {Chen}}, \bibinfo {author} {\bibfnamefont
  {X.}~\bibnamefont {Wu}}, \bibinfo {author} {\bibfnamefont {W.}~\bibnamefont {Jiang}}, \bibinfo {author} {\bibfnamefont {E.}~\bibnamefont {Berger}}, \bibinfo {author} {\bibfnamefont {Y.}~\bibnamefont {Li}}, \bibinfo {author} {\bibfnamefont {S.}~\bibnamefont {Chen}}, \bibinfo {author} {\bibfnamefont {A.~J.}\ \bibnamefont {Gabourie}}, \bibinfo {author} {\bibfnamefont {H.}~\bibnamefont {Dong}}, \bibinfo {author} {\bibfnamefont {S.}~\bibnamefont {Xiong}}, \bibinfo {author} {\bibfnamefont {N.}~\bibnamefont {Wei}}, \bibinfo {author} {\bibfnamefont {Y.}~\bibnamefont {Chen}}, \bibinfo {author} {\bibfnamefont {J.}~\bibnamefont {Xu}}, \bibinfo {author} {\bibfnamefont {F.}~\bibnamefont {Ding}}, \bibinfo {author} {\bibfnamefont {Z.}~\bibnamefont {Sun}}, \bibinfo {author} {\bibfnamefont {T.}~\bibnamefont {Ala-Nissila}}, \bibinfo {author} {\bibfnamefont {A.}~\bibnamefont {Harju}}, \bibinfo {author} {\bibfnamefont {J.}~\bibnamefont {Zheng}}, \bibinfo {author} {\bibfnamefont {P.}~\bibnamefont {Guan}}, \bibinfo {author}
  {\bibfnamefont {P.}~\bibnamefont {Erhart}}, \bibinfo {author} {\bibfnamefont {J.}~\bibnamefont {Sun}}, \bibinfo {author} {\bibfnamefont {W.}~\bibnamefont {Ouyang}}, \bibinfo {author} {\bibfnamefont {Y.}~\bibnamefont {Su}},\ and\ \bibinfo {author} {\bibfnamefont {Z.}~\bibnamefont {Fan}},\ }\href {https://doi.org/doi: 10.1002/mgea.70028} {\bibfield  {journal} {\bibinfo  {journal} {Materials Genome Engineering Advances}\ }\textbf {\bibinfo {volume} {3}},\ \bibinfo {pages} {e70028} (\bibinfo {year} {2025})}\BibitemShut {NoStop}%
\bibitem [{\citenamefont {Lindgren}\ \emph {et~al.}(2024)\citenamefont {Lindgren}, \citenamefont {Rahm}, \citenamefont {Fransson}, \citenamefont {Eriksson}, \citenamefont {{\"O}sterbacka}, \citenamefont {Fan},\ and\ \citenamefont {Erhart}}]{LinRahFra24}%
  \BibitemOpen
  \bibfield  {author} {\bibinfo {author} {\bibfnamefont {E.}~\bibnamefont {Lindgren}}, \bibinfo {author} {\bibfnamefont {M.}~\bibnamefont {Rahm}}, \bibinfo {author} {\bibfnamefont {E.}~\bibnamefont {Fransson}}, \bibinfo {author} {\bibfnamefont {F.}~\bibnamefont {Eriksson}}, \bibinfo {author} {\bibfnamefont {N.}~\bibnamefont {{\"O}sterbacka}}, \bibinfo {author} {\bibfnamefont {Z.}~\bibnamefont {Fan}},\ and\ \bibinfo {author} {\bibfnamefont {P.}~\bibnamefont {Erhart}},\ }\href {https://doi.org/10.21105/joss.06264} {\bibfield  {journal} {\bibinfo  {journal} {Journal of Open Source Software}\ }\textbf {\bibinfo {volume} {9}},\ \bibinfo {pages} {6264} (\bibinfo {year} {2024})}\BibitemShut {NoStop}%
\bibitem [{\citenamefont {Schaul}\ \emph {et~al.}(2011)\citenamefont {Schaul}, \citenamefont {Glasmachers},\ and\ \citenamefont {Schmidhuber}}]{schaul2011high}%
  \BibitemOpen
  \bibfield  {author} {\bibinfo {author} {\bibfnamefont {T.}~\bibnamefont {Schaul}}, \bibinfo {author} {\bibfnamefont {T.}~\bibnamefont {Glasmachers}},\ and\ \bibinfo {author} {\bibfnamefont {J.}~\bibnamefont {Schmidhuber}},\ }in\ \href {https://doi.org/10.1145/2001576.2001692} {\emph {\bibinfo {booktitle} {Proceedings of the 13th Annual Conference on Genetic and Evolutionary Computation}}},\ \bibinfo {series and number} {GECCO '11}\ (\bibinfo  {publisher} {Association for Computing Machinery},\ \bibinfo {address} {New York, NY, USA},\ \bibinfo {year} {2011})\ p.\ \bibinfo {pages} {845–852}\BibitemShut {NoStop}%
\bibitem [{\citenamefont {Kresse}\ and\ \citenamefont {Furthm\"uller}(1996)}]{KreFur1996-1}%
  \BibitemOpen
  \bibfield  {author} {\bibinfo {author} {\bibfnamefont {G.}~\bibnamefont {Kresse}}\ and\ \bibinfo {author} {\bibfnamefont {J.}~\bibnamefont {Furthm\"uller}},\ }\href {https://doi.org/10.1103/PhysRevB.54.11169} {\bibfield  {journal} {\bibinfo  {journal} {Physical Review B}\ }\textbf {\bibinfo {volume} {54}},\ \bibinfo {pages} {11169} (\bibinfo {year} {1996})}\BibitemShut {NoStop}%
\bibitem [{\citenamefont {Bl\"ochl}(1994)}]{Blo94}%
  \BibitemOpen
  \bibfield  {author} {\bibinfo {author} {\bibfnamefont {P.~E.}\ \bibnamefont {Bl\"ochl}},\ }\href {https://doi.org/10.1103/PhysRevB.50.17953} {\bibfield  {journal} {\bibinfo  {journal} {Physical Review B}\ }\textbf {\bibinfo {volume} {50}},\ \bibinfo {pages} {17953} (\bibinfo {year} {1994})}\BibitemShut {NoStop}%
\bibitem [{\citenamefont {Kresse}\ and\ \citenamefont {Joubert}(1999)}]{KreJou99}%
  \BibitemOpen
  \bibfield  {author} {\bibinfo {author} {\bibfnamefont {G.}~\bibnamefont {Kresse}}\ and\ \bibinfo {author} {\bibfnamefont {D.}~\bibnamefont {Joubert}},\ }\href {https://doi.org/10.1103/PhysRevB.59.1758} {\bibfield  {journal} {\bibinfo  {journal} {Physical Review B}\ }\textbf {\bibinfo {volume} {59}},\ \bibinfo {pages} {1758} (\bibinfo {year} {1999})}\BibitemShut {NoStop}%
\bibitem [{\citenamefont {Dion}\ \emph {et~al.}(2004)\citenamefont {Dion}, \citenamefont {Rydberg}, \citenamefont {Schr\"oder}, \citenamefont {Langreth},\ and\ \citenamefont {Lundqvist}}]{DioRydSch04}%
  \BibitemOpen
  \bibfield  {author} {\bibinfo {author} {\bibfnamefont {M.}~\bibnamefont {Dion}}, \bibinfo {author} {\bibfnamefont {H.}~\bibnamefont {Rydberg}}, \bibinfo {author} {\bibfnamefont {E.}~\bibnamefont {Schr\"oder}}, \bibinfo {author} {\bibfnamefont {D.~C.}\ \bibnamefont {Langreth}},\ and\ \bibinfo {author} {\bibfnamefont {B.~I.}\ \bibnamefont {Lundqvist}},\ }\href {https://doi.org/10.1103/PhysRevLett.92.246401} {\bibfield  {journal} {\bibinfo  {journal} {Physical Review Letters}\ }\textbf {\bibinfo {volume} {92}},\ \bibinfo {pages} {246401} (\bibinfo {year} {2004})}\BibitemShut {NoStop}%
\bibitem [{\citenamefont {Berland}\ and\ \citenamefont {Hyldgaard}(2014)}]{BerHyl2014}%
  \BibitemOpen
  \bibfield  {author} {\bibinfo {author} {\bibfnamefont {K.}~\bibnamefont {Berland}}\ and\ \bibinfo {author} {\bibfnamefont {P.}~\bibnamefont {Hyldgaard}},\ }\href {https://doi.org/10.1103/PhysRevB.89.035412} {\bibfield  {journal} {\bibinfo  {journal} {Physical Review B}\ }\textbf {\bibinfo {volume} {89}},\ \bibinfo {pages} {035412} (\bibinfo {year} {2014})}\BibitemShut {NoStop}%
\bibitem [{\citenamefont {Gonze}\ and\ \citenamefont {Lee}(1997)}]{Gonze1997}%
  \BibitemOpen
  \bibfield  {author} {\bibinfo {author} {\bibfnamefont {X.}~\bibnamefont {Gonze}}\ and\ \bibinfo {author} {\bibfnamefont {C.}~\bibnamefont {Lee}},\ }\href {https://doi.org/10.1103/PhysRevB.55.10355} {\bibfield  {journal} {\bibinfo  {journal} {Physical Review B}\ }\textbf {\bibinfo {volume} {55}},\ \bibinfo {pages} {10355} (\bibinfo {year} {1997})}\BibitemShut {NoStop}%
\bibitem [{\citenamefont {Sun}\ \emph {et~al.}(2010)\citenamefont {Sun}, \citenamefont {Shen},\ and\ \citenamefont {Allen}}]{SunSheAll2010}%
  \BibitemOpen
  \bibfield  {author} {\bibinfo {author} {\bibfnamefont {T.}~\bibnamefont {Sun}}, \bibinfo {author} {\bibfnamefont {X.}~\bibnamefont {Shen}},\ and\ \bibinfo {author} {\bibfnamefont {P.~B.}\ \bibnamefont {Allen}},\ }\href {https://doi.org/10.1103/PhysRevB.82.224304} {\bibfield  {journal} {\bibinfo  {journal} {Physical Review B}\ }\textbf {\bibinfo {volume} {82}},\ \bibinfo {pages} {224304} (\bibinfo {year} {2010})}\BibitemShut {NoStop}%
\bibitem [{\citenamefont {Rohskopf}\ \emph {et~al.}(2022)\citenamefont {Rohskopf}, \citenamefont {Li}, \citenamefont {Luo},\ and\ \citenamefont {Henry}}]{Rohskopf2022}%
  \BibitemOpen
  \bibfield  {author} {\bibinfo {author} {\bibfnamefont {A.}~\bibnamefont {Rohskopf}}, \bibinfo {author} {\bibfnamefont {R.}~\bibnamefont {Li}}, \bibinfo {author} {\bibfnamefont {T.}~\bibnamefont {Luo}},\ and\ \bibinfo {author} {\bibfnamefont {A.}~\bibnamefont {Henry}},\ }\href {https://doi.org/10.1088/1361-651X/ac5ebb} {\bibfield  {journal} {\bibinfo  {journal} {Modelling and Simulation in Materials Science and Engineering}\ }\textbf {\bibinfo {volume} {30}},\ \bibinfo {pages} {045010} (\bibinfo {year} {2022})}\BibitemShut {NoStop}%
\bibitem [{\citenamefont {Berger}\ \emph {et~al.}(2025)\citenamefont {Berger}, \citenamefont {Fransson}, \citenamefont {Eriksson}, \citenamefont {Lindgren}, \citenamefont {Wahnström}, \citenamefont {Rod},\ and\ \citenamefont {Erhart}}]{BerFraEri25}%
  \BibitemOpen
  \bibfield  {author} {\bibinfo {author} {\bibfnamefont {E.}~\bibnamefont {Berger}}, \bibinfo {author} {\bibfnamefont {E.}~\bibnamefont {Fransson}}, \bibinfo {author} {\bibfnamefont {F.}~\bibnamefont {Eriksson}}, \bibinfo {author} {\bibfnamefont {E.}~\bibnamefont {Lindgren}}, \bibinfo {author} {\bibfnamefont {G.}~\bibnamefont {Wahnström}}, \bibinfo {author} {\bibfnamefont {T.~H.}\ \bibnamefont {Rod}},\ and\ \bibinfo {author} {\bibfnamefont {P.}~\bibnamefont {Erhart}},\ }\href {https://doi.org/10.1016/j.cpc.2025.109759} {\bibfield  {journal} {\bibinfo  {journal} {Computer Physics Communications}\ }\textbf {\bibinfo {volume} {316}},\ \bibinfo {pages} {109759} (\bibinfo {year} {2025})}\BibitemShut {NoStop}%
\bibitem [{\citenamefont {Kemeny}\ \emph {et~al.}(1986)\citenamefont {Kemeny}, \citenamefont {Mahanti},\ and\ \citenamefont {Kaplan}}]{Kemeny1986}%
  \BibitemOpen
  \bibfield  {author} {\bibinfo {author} {\bibfnamefont {G.}~\bibnamefont {Kemeny}}, \bibinfo {author} {\bibfnamefont {S.~D.}\ \bibnamefont {Mahanti}},\ and\ \bibinfo {author} {\bibfnamefont {T.~A.}\ \bibnamefont {Kaplan}},\ }\href {https://doi.org/10.1103/PhysRevB.34.6288} {\bibfield  {journal} {\bibinfo  {journal} {Physical Review B}\ }\textbf {\bibinfo {volume} {34}},\ \bibinfo {pages} {6288} (\bibinfo {year} {1986})}\BibitemShut {NoStop}%
\bibitem [{\citenamefont {Dabelow}\ \emph {et~al.}(2019)\citenamefont {Dabelow}, \citenamefont {Bo},\ and\ \citenamefont {Eichhorn}}]{dabelow2019}%
  \BibitemOpen
  \bibfield  {author} {\bibinfo {author} {\bibfnamefont {L.}~\bibnamefont {Dabelow}}, \bibinfo {author} {\bibfnamefont {S.}~\bibnamefont {Bo}},\ and\ \bibinfo {author} {\bibfnamefont {R.}~\bibnamefont {Eichhorn}},\ }\href {https://doi.org/10.1103/PhysRevX.9.021009} {\bibfield  {journal} {\bibinfo  {journal} {Physical Review X}\ }\textbf {\bibinfo {volume} {9}},\ \bibinfo {pages} {021009} (\bibinfo {year} {2019})}\BibitemShut {NoStop}%
\end{thebibliography}
\end{document}